\documentclass[paper]{JHEP3}

\usepackage{epsfig,multicol,bbm}
\usepackage{amsmath}
\usepackage{verbatim}
\usepackage{graphicx}
\usepackage{psfrag}
\usepackage{epstopdf}
\usepackage{epspdfconversion}
\usepackage{amssymb}

\newcommand\fverb{\setbox\fverbbox=\hbox\bgroup\verb}
\newcommand\fverbdo{\egroup\medskip\noindent%
            \fbox{\unhbox\fverbbox}\ }
\newcommand\fverbit{\egroup\item[\fbox{\unhbox\fverbbox}]}
\newbox\fverbbox

\title{Parameterizing scalar-tensor theories for cosmological probes}
\author{$^{1,2}$Stephen A Appleby, $^{1,2,3}$Jochen Weller \\
$^{1}$Excellence Cluster Universe, Boltzmannstr. 2, 85748 Garching, Germany \\
$^{2}$University Observatory, Ludwig-Maximillians University Munich,  \\ Scheinerstr. 1, 81679 Munich, Germany \\ 
$^{3}$Max-Planck-Institut f\"{u}r extraterrestrische Physik, \\ Giessenbachstrasse, 85748 Garching, Germany \\
Email: \email{stephen.appleby@ph.tum.de}, \email{jochen.weller@usm.lmu.de}}

\date{\today}

\abstract{We study the evolution of density perturbations for a class of $f(R)$ models which closely mimic $\Lambda$CDM background cosmology. Using the quasi-static approximation, and the fact that these models are equivalent to scalar-tensor gravity, we write the modified Friedmann and cosmological perturbation equations in terms of the mass $M$ of the scalar field. Using the perturbation equations, we then derive an analytic expression for the growth parameter $\gamma$ in terms of $M$, and use our result to reconstruct the linear matter power spectrum. We find that the power spectrum at $z \sim 0$ is characterized by a tilt relative to its General Relativistic form, with increased power on small scales. We discuss how one has to modify the standard, constant $\gamma$ prescription in order to study structure formation for this class of models. Since $\gamma$ is now scale and time dependent, both the amplitude and transfer function associated with the linear matter power spectrum will be modified. We suggest a simple parameterization for the mass of the scalar field, which allows us to calculate the matter power spectrum for a broad class of $f(R)$ models.}

\begin{document}

\section{Introduction}

Cosmological data, arising from anisotropies of the cosmic microwave background \cite{Spergel:2006hy,Komatsu:2008hk}, large scale structure \cite{Eisenstein:2005su} and measurements of type Ia supernovae \cite{Riess:1998cb,Perlmutter:1998np}, indicate that the expansion of the Universe is currently accelerating. The simplest approach to modeling this epoch is to postulate the existence of a very small but non-zero vacuum energy, which can dominate at late times and accelerate the expansion by virtue of having negative pressure (specifically, an equation of state parameter $w = -1$.) However, the extreme fine tuning implicit in such a model has driven a search for alternative dark energy candidates, where the acceleration is attributed to one or more dynamical fields \cite{Copeland:2006wr}, (alternative approaches include higher dimensional physics \cite{Dvali:2000hr,Deffayet:2000uy,Deffayet:2001pu} or the backreaction of inhomogeneities on an isotropic, averaged metric; see \cite{Buchert:2006rq} for a recent discussion.)

Rather than introducing additional matter content to the Universe, an alternative approach to modeling the current epoch is to postulate that at cosmological distance scales, gravity deviates from its standard General Relativistic description. This can be achieved, for example, by introducing fields which mediate gravity in addition to the standard spin-2 graviton. In this work we consider the $f(R)$ subclass of scalar-tensor theories, described by the action

\begin{equation} \label{eq:1} S = \int \sqrt{-g} d^{4}x \left[ { R + f(R) \over 16\pi G} + {\cal L}_{\rm m}\right] ,\end{equation}

\noindent where ${\cal L}_{\rm m}$ is the Lagrange density of standard matter and radiation, and $f(R)$ is an unspecified function of the Ricci scalar \footnote {The sign conventions in this paper are:
the metric signature (-+++), the curvature tensor $R^{\sigma}_{\
\mu\rho\nu} = \partial_{\rho}\Gamma^{\sigma}_{\mu\nu} - ...,~R_{\mu\nu}=R^{\sigma}_{\
\mu\sigma\nu}$, so that the Ricci scalar
$R=R_{\mu}^{\mu}>0$ for the de Sitter space-time and the
matter-dominated cosmological epoch. In addition, we set $c = 1$ throughout.}. These models contain an additional scalar field in the gravitational particle sector, which we will call the scalaron. The scalaron has the unique property that its mass depends on the background curvature of spacetime $M = M(R_{\rm background})$; this fact proves important in evading solar system tests of gravity. A large body of literature has been devoted to models of the form ($\ref{eq:1}$), and we direct the reader to \cite{Sotiriou:2008rp,DeFelice:2010aj,Nojiri:2006ri,Capozziello:2007ec} and references therein for a detailed recent review.

In principle one could use an arbitrary $f(R)$ function in the action ($\ref{eq:1}$), however in order to be both observationally and theoretically viable these models must respect a large list of consistency requirements. To begin,  the $f(R)$ function must satisfy $1+f_{\rm R} > 0$ and $f_{\rm RR} > 0$ (throughout the paper, $R$ subscripts denote derivatives with respect to $R$) in all dynamically accessible regions to cosmology, to evade ghost and other instabilities \cite{st1,H73,GS79,RR70,d1}. 

$f(R)$ models must also have a viable Newtonian limit. The conditions required for an $f(R)$ function to reproduce Newtonian gravity at a particular curvature scale $R_{\rm N}$ are

\begin{equation} |f(R_{\rm N})|  \ll R_{\rm N} \qquad |f_{\rm R}| \ll 1 \qquad R_{\rm N}f_{RR} \ll 1 . \end{equation}

\noindent These conditions must hold, for example, in the Solar System, where gravity is accurately described by weak field Newtonian physics.

Finally, prospective $f(R)$ models must be able to reproduce a viable cosmology \cite{at10,am1,Nojiri:2006gh}, with a period of radiation domination followed by a standard matter era where the scale factor evolves approximately as $a(t) \sim t^{2/3}$. It has been shown that only a very restricted subclass of $f(R)$ models can reproduce an expansion history consistent with observations, as the scale factor typically evolves as $a(t) \propto t^{1/2}$ during matter domination \cite{at10,am1}.

In addition to the above theoretical considerations, Solar system and laboratory tests of gravity provide stringent observational constraints on $f(R)$ models \cite{Hu:2007nk},\cite{Chiba:2003ir},\cite{Tsujikawa:2008uc}. For example the Cassini probe measurement of the parameterized post Newtonian (PPN) parameter $\gamma_{\rm PPN}$ in the solar system \cite{Bertotti:2003rm} can be translated into a constraint on $ |f_{R}(R_{\rm g})|$ \cite{Hu:2007nk}

\begin{equation} \label{eq:p5} |f_{R}(R_{\rm g})| < 4.9 \times 10^{-11} , \end{equation}

\noindent where $R_{\rm g} \sim 8\pi G \rho_{\rm g}$ and $\rho_{\rm g} \simeq 10^{-24}$ g ${\rm cm^{-3}}$ is the typical galactic background energy density. The condition ($\ref{eq:p5}$)  might appear to be a particularly strong condition on $f(R)$ models ($f_{R} = 0$ is the General Relativistic limit), however the `background' dependence of the mass of the scalar field allows us to construct $f(R)$ models that easily evade this bound. Viable models are constructed such that the (squared) mass of the scalar field is much larger than the local curvature, $M^{2}(R_{\rm g}) \gg R_{\rm g}$. In this case, the scalar field will not propagate on macroscopic distance scales, and deviations from GR will be suppressed in the solar system.

More recently it has been argued that stringent observational constraints on $f(R)$ gravity arise by considering structure formation, as the evolution of density perturbations will be significantly modified by the scalar field. The most commonly studied model in this regard is the `minimal' $f(R)$ model \cite{Song:2006ej}, so called because it has an expansion history that is identically $\Lambda$CDM by construction (although see \cite{Fay:2007uy,Dunsby:2010wg}), and possesses only one additional parameter as compared to the standard cosmological model. This additional parameter is taken to be $B_{0}$, defined as $B_{0} \equiv B(a=1)$ where the function $B(a)$ is given by

\begin{equation} B= {f_{\rm RR} \over 1+ f_{\rm R}} {R' H \over H'} ,\end{equation}

\noindent where primes denote derivatives with respect to $\ln(a)$. In a series of papers \cite{Hu:2007pj}\cite{Song:2007da}\cite{Lombriser:2010mp}, it has been concluded that current cosmological data sets impose \cite{Lombriser:2010mp} $B_{0} < 1.1 \times 10^{-3}$ at $95\%$ C.L \footnote{see also \cite{Schmidt:2009am},\cite{Smith:2009fn},\cite{Martinelli:2009ek},\cite{Nojiri:2006be} for constraints on $f(R)$ models} ($B_{0} = 0$ is the General Relativistic limit.)

Observational tests of $f(R)$ models place lower limits on the mass of the scalar field. If the mass is much larger than the background curvature, the scalaron will essentially be non-dynamical, and this class of models will mimic General Relativity. In the solar system, the constraint ($\ref{eq:p5}$) implies that the scalaron mass must be very large; we would like to consider the extent to which the mass is allowed to relax on cosmological scales. It is the aim of this work to construct a parameterization of $f(R)$ models, which describes both the expansion and growth histories, in terms of the scalar field mass $M(a)$. We derive all results in a model independent manner (all quantities will be written in terms of $M(a)$), however we will sometimes resort to specific models when we wish to numerically evolve the field equations. We will use the following two functions

\begin{eqnarray} \label{eq:p1} & & f(R) = -{R_{\rm vac} \over 2} + 2g \epsilon e^{-2(R/\epsilon - b)}  + {R^{2} \over 6M_{0}^{2}} , \\ \label{eq:p2} & & f(R) =       - {R_{\rm vac} \over 2} + \lambda R_{\rm vac} \left( {R_{\rm vac} \over R}\right)^{2n}  + {R^{2} \over 6M_{0}^{2}} , \end{eqnarray}

\noindent  where $\epsilon = R_{\rm vac}/2g(b+\log[2\cosh[b]])$, $g,n,\lambda,b$ are model parameters, we will take $M_{0}$ to be an inflationary mass scale and $R_{\rm vac} = 12H_{0}^{2}\Omega_{\Lambda}$ is the curvature associated with the vacuum. These two functional forms, which will be referred to as the exponential and power law models respectively, represent expansions around General Relativity that are commonly studied in the literature \cite{Hu:2007nk},\cite{st},\cite{ap},\cite{ts10},\cite{Cognola:2007zu}\cite{Linder:2009jz},\cite{Bamba:2010ws}. We stress that these functions are not globally viable $f(R)$ models, however they are sufficient for our purposes in the sense that they are free from instabilities, and reproduce the standard cosmology, over redshifts $z = (10^{3},0)$.

The paper will proceed as follows. We begin by studying the growth parameter $\gamma$. In section \ref{sec:q1} we discuss $\gamma$ as a parameter in General Relativity. In section \ref{sec:q2} and \ref{sec:q3} we consider the field equations for $f(R)$ models, and how they can be simplified using the `quasi-static' approximation. We use this approximation scheme to construct a model independent expression for $\gamma$ in terms of the mass of the scalar field $M(a)$. In section \ref{sec:i7} we introduce a simple yet representative parameterization of $M(a)$ and discuss how the matter power spectrum depends on this function. We end with a discussion on the use of $\gamma$ as a parameterization for growth, and how our approach relates to existing work in the literature.

\section{\label{sec:2}The growth parameter}

We begin by studying the growth parameter $\gamma$ \cite{peebles:1980,Lahav:1991wc,Wang:1998gt,Linder:2005in,Linder:2007hg}, defined as

\begin{equation} \label{eq:2}\gamma \equiv {\log [d\log[\delta_{\rm m}] /d\log[a] ] \over \log[\Omega_{\rm m}]} , \end{equation}

\noindent where `$m$' subscripts denote the total matter component, $\delta_{\rm m} \equiv \delta \rho_{\rm m} /\rho_{\rm m}$ and $\Omega_{\rm m}$ is defined in the usual manner,

\begin{equation} \label{eq:3} \Omega_{\rm m} = {8\pi G \rho_{\rm m} \over 3H^{2}} .\end{equation}

\subsection{\label{sec:q1}The growth parameter in General Relativity}

\noindent  We first briefly re-derive the value of $\gamma$ in a standard $\Lambda$CDM cosmology; $\gamma \simeq 6/11$ over the redshift range $z \sim (0,10)$. To obtain this result, consider the evolution of density perturbations in a cosmological background containing two perfect fluids; pressureless matter (which will dominate for $z \sim (1,10)$) and a subdominant component with unspecified equation of state $P = \bar{w}\rho$ (throughout this paper we will make the simplifying assumption that the Universe is spatially flat, $\Omega_{\rm k0} = 0$). Considering only the evolution of matter perturbations in this epoch, $\delta_{\rm m}$ evolves according to

\begin{equation} \label{eq:4} a^{2}\delta''_{\rm m} + \left[ 3 - {3 \over 2}\Omega_{\rm m} - {3 \over 2}(1+\bar{w})(1 - \Omega_{\rm m}) \right] a \delta'_{\rm m}  - {3 \over 2} \Omega_{\rm m}\delta_{\rm m} = 0,  \end{equation}

\noindent where throughout the paper primes denote differentiation with respect to $a$. By differentiating the definition of $\gamma$ ($\ref{eq:2}$) with respect to $a$, and using ($\ref{eq:4}$) to remove second derivatives of $\delta_{\rm m}$, we obtain an equation describing the evolution of $\gamma$,

\begin{eqnarray}\nonumber  \gamma' = & &  {1 \over \Omega_{\rm m}^{\gamma}\log[\Omega_{\rm m}]} \left( {3 \over 2}{\Omega_{\rm m} \over a} + \left({3\Omega_{\rm m} \over 2} - 2\right) {\Omega_{\rm m}^{\gamma} \over a} - {\Omega_{\rm m}^{2\gamma} \over a} + {3 \over 2}(1+\bar{w})(1 - \Omega_{\rm m}){\Omega_{\rm m}^{\gamma} \over a}\right) \\ \label{eq:5} & & \quad - {3 \bar{w} \over a}{\gamma \over  \log[\Omega_{\rm m}]}(1 - \Omega_{\rm m}) .\end{eqnarray}

\noindent To solve ($\ref{eq:5}$), we use the fact that for $z =(1,10)$, we can expand $\Omega_{\rm m}$ as

\begin{equation} \Omega_{\rm m} = 1 - {\Omega_{\rm \bar{w}0} \over \Omega_{\rm m0}}a^{-3\bar{w}} + {\cal O}\left(   {\Omega_{\rm \bar{w}0}^{2} \over \Omega_{\rm m0}^{2}}a^{-6\bar{w}}  \right)  ,  \end{equation}

\noindent  where $\Omega_{\rm m0}$ and $\Omega_{\rm \bar{w}0}$ are, respectively, the fractional densities of the matter and subdominant fluid at the present time. Expanding equation ($\ref{eq:5}$), we obtain a series solution 

\begin{equation} \gamma = {3(1-\bar{w}) \over 5 - 6\bar{w}} + {\cal O} \left( { \Omega_{\rm \bar{w}0} \over \Omega_{\rm m0}} a^{-3\bar{w}} \right) . \end{equation}

\noindent At first order in the expansion $\gamma$ is sensitive only to the equation of state $\bar{w}$ of the sub-dominant energy component. For $z = (1,10)$, we take $\bar{w} = -1$ and find $\gamma \simeq  6/11$ as expected. 

At low redshift $z < 1$, we can no longer use the expansion $a \ll 1$ to derive an expression for $\gamma$. However, it is straightforward to show that  for a $\Lambda$CDM cosmology, $\gamma$ will asymptote to a value $\gamma \to 2/3$ for $a \gg 1$. This asymptotic behaviour is logarithmic with respect to the scale factor, and hence we can assume that $\gamma$ will not deviate significantly from the value $\gamma \simeq 6/11$ for $ 0 < z < 1$. At high redshift $z > 10$, constructing an analytic form for $\gamma$ is complicated by the fact that we must take into account the radiation component in both the perturbation equations and $\Omega_{\rm m}(a)$ (since $\rho_{\rm r} > \rho_{\Lambda}$ for $z > 10$, where $\rho_{\rm r}$ is the radiation energy density.)

For dark energy models with a constant equation of state, $\gamma$ is an excellent parameterization for the growth \cite{Linder:2005in},\cite{Linder:2007hg}. For redshifts of interest to structure formation, $\gamma$ is approximately constant, $a\gamma' \ll \gamma$, and is only weakly dependent on cosmological parameters such as $\Omega_{m0}$. In addition, knowledge of $\gamma$ is sufficient to reconstruct the time evolution of the matter power spectrum for $z < 10$. This is achieved by integrating the expression

\begin{equation} \label{eq:j1} \delta_{\rm m} = \exp \left[ \int \Omega_{\rm m}^{\gamma} d\ln a \right] . \end{equation}

\subsection{\label{sec:q2}The growth parameter in $f(R)$ models}

The above results are well known in General Relativity \cite{peebles:1980,Lahav:1991wc,Wang:1998gt,Linder:2005in}. We now perform a similar calculation for $f(R)$ models. We note that the growth parameter in the context of modified gravity has been considered in a number of recent works \cite{Brax:2009ab,Motohashi:2010qj,Motohashi:2010tb,Narikawa:2009ux} (see also \cite{Linder:2007hg} for an earlier treatment of $\gamma$ in modified gravity models.)

 At zeroth order in the perturbations, we have

\begin{eqnarray}\label{eq:i5} & & 3 H^{2} = 8\pi G (\rho_{\rm m} + \rho_{\rm r})+ {f_{\rm R}R - f \over 2} - 3H \dot{f}_{\rm R} , \\ \label{eq:i6} & & -2\dot{H} = 8\pi G \rho_{\rm m} + {32\pi G \rho_{\rm r} \over 3} + \ddot{f}_{\rm R} - H \dot{f}_{\rm R} ,  \end{eqnarray}

\noindent where dots denote derivatives with respect to time and $\rho_{m},\rho_{r}$ are the energy densities of matter and radiation respectively. The $f(R)$ terms in ($\ref{eq:i5},\ref{eq:i6}$) are often written as components of a perfect fluid,

\begin{eqnarray} \label{eq:a1} & & 3 H^{2} = 8\pi G (\rho_{\rm m} + \rho_{\rm r} + \rho_{\rm f}) , \\ \label{eq:a2} & & -2\dot{H} = 8\pi G \rho_{\rm m} + {32 \pi G \rho_{\rm r} \over 3} + 8\pi G(\rho_{\rm f} + P_{\rm f}) , \end{eqnarray}

\noindent where $\rho_{\rm f}$ and $P_{\rm f}$ are effective density and pressure terms

\begin{eqnarray} & & \label{eq:f1} 8\pi G \rho_{\rm f} = {f_{\rm R}R - f \over 2} - 3H\dot{f}_{\rm R} , \\ & & \label{eq:f2} 8\pi G P_{\rm f} = -{(Rf_{\rm R} - f)\over 2} + \ddot{f}_{\rm  R} + 2H\dot{f}_{\rm R} .\end{eqnarray}

\noindent The equation of state parameter $w_{f}$ for this fluid is given by

\begin{equation} \label{eq:w1} w_{\rm f} = {2\ddot{f} _{\rm R}+ 4H\dot{f}_{\rm R} - (f_{\rm R}R - f) \over -6H\dot{f}_{\rm R} + (f_{\rm R}R - f)} ,\end{equation}

\noindent  from which it is clear that the modified gravity terms will mimic dark energy with equation of state $w_{\rm f} \simeq -1$ whenever $f_{\rm R}$ is approximately static; $H\dot{f}_{\rm R}, \ddot{f}_{\rm R} \ll Rf_{\rm R} - f$. Although we have rewritten the modified gravity terms as $\rho_{\rm f}$ and $P_{f}$, we stress that these quantities are functions of $\dot{H},\ddot{H}, \dddot{H}$, and the modified gravity equations are fourth order in derivatives of the scale factor.

We are interested in the behaviour of the density perturbations $\delta_{\rm m}$ in this class of models. The equations describing the evolution of scalar perturbations in f(R) models have been derived in \cite{Bean:2006up}. In the Newtonian gauge, using sign conventions such that the metric potentials are defined as

\begin{equation} ds^{2} = -(1+2\psi)dt^{2} + a^{2}(t)(1-2\phi)\gamma_{ij}dx^{i}dx^{j} , \end{equation}

\noindent we find \cite{Bean:2006up}

\begin{eqnarray} \label{eq:10} & & \ddot{\delta}_{\rm m} + 2H\dot{\delta}_{\rm m} + k^{2}\psi - 3\ddot{\phi} - 6H\dot{\phi}  = 0 , \\ \label{eq:11} & & (1+f_{\rm R})(\psi - \phi) + f_{\rm RR} \epsilon  = 0 ,   \\  \nonumber & & (1+f_{\rm R}) \left[ 2{k^{2} \over a^{2}} \phi +   6H(\dot{\phi} + H\psi)\right] + 3f_{\rm RR} (\dot{H} + H^{2}) -  \left({k^{2} \over a^{2}}f_{\rm RR} + 3H\dot{f}_{\rm RR}\right) \epsilon  + \\ \label{eq:12} & &  - 3Hf_{\rm RR}  \dot{\epsilon} + \dot{f}_{\rm R}(6H\psi + 3\dot{\phi})  = -8\pi G  \rho_{\rm m}\delta_{\rm m} ,  \\ \label{eq:15} & &  \epsilon \equiv -12(\dot{H} + 2H^{2})\psi - 6H\dot{\psi} + 2{k^{2} \over a^{2}} \psi - 6(\ddot{\phi} + 4H\dot{\phi}) - 4{k^{2} \over a^{2}}\phi  , \end{eqnarray}

\noindent where we have neglected perturbations in the radiation component (this is a valid approximation at late times.)

The background and perturbation equations ($\ref{eq:i5},\ref{eq:i6},\ref{eq:10}-\ref{eq:15}$) comprise a complicated system of coupled, fourth order differential equations, and solving them (even numerically) is a highly non-trivial task. However, for the class of so-called `viable' models which can be written as expansions around General Relativity (see for example ($\ref{eq:p1},\ref{eq:p2}$)), we can use the quasi-static approximation \cite{Boisseau:2000pr},\cite{Zhang:2005vt} to simplify the dynamics.

\subsection{\label{sec:q3}Field equations in the quasi-static approximation}

In a cosmological context, the quasi-static approximation posits that the mass of the scalar field remains larger than the Hubble parameter at all redshifts; $H(a)/ M(a) \ll 1$. In this case, the scalar field will only be weakly dynamical throughout the cosmological history and the background evolution will closely mimic $\Lambda$CDM, with corrections of order $H^{2} /M^{2}$. 

To see what happens to the modified Friedmann equation ($\ref{eq:a1}$) in the quasi-static limit, we solve the trace of the gravitational field  equations

\begin{equation}\label{eq:e1} 3f_{\rm RR}\ddot{R} + 3f_{\rm RRR}(\dot{R})^{2} + 9H f_{\rm RR} \dot{R} + R - Rf_{\rm R} + 2f = - 8\pi G T .\end{equation}

\noindent For models that reduce to expansions around General Relativity (such as ($\ref{eq:p1},\ref{eq:p2}$)), we can expand the function $f(R)$ around a cosmological constant, $f(R) =  - R_{\rm vac}/2 + \bar{f}(R)$ \footnote{It is important to stress that there is no true cosmological constant in these models; they generically admit Minkowski space as a solution to the vacuum field equations. However, for the region of interest to cosmology we can expand the $f(R)$ function around a pseudo constant $R_{\rm vac}/2$.}, where $\bar{f}$ is an unspecified function of the Ricci scalar. Doing so, it has been found \cite{st} that equation ($\ref{eq:e1}$) has approximate solution $R \simeq R_{\rm GR} + \delta R_{\rm ind} + \delta R_{\rm osc}$, where $R_{\rm GR}$ and $\delta R_{\rm ind}$ are given by

\small

\begin{eqnarray} & & {R_{\rm GR} \over 12 H_{0}^{2}} =   {\Omega_{\rm m0} \over 4a^{3}} + \Omega_{\Lambda}  ,  \\ \label{eq:y1} & & \delta R_{\rm ind} =  -3\bar{f}_{\rm RR}\ddot{R}_{\rm GR} -3\bar{f}_{\rm RRR}(\dot{R}_{\rm GR})^{2}   - 9H \bar{f}_{\rm RR} \dot{R}_{\rm GR} + R_{\rm GR}\bar{f}_{\rm R} - 2\bar{f} . \end{eqnarray}

\normalsize

\noindent In ($\ref{eq:y1}$) and unless stated otherwise throughout the paper, all $\bar{f},\bar{f}_{\rm R},\bar{f}_{\rm RR},\bar{f}_{\rm RRR}$ terms are functions of $R_{\rm GR}$; $\bar{f} = \bar{f}(R_{\rm GR})$. 

$\delta R_{\rm osc}$ is a component of the Ricci scalar which oscillates with high frequency $\omega \sim M(a)$. The oscillating component has been discussed in previous works, see for example \cite{st,Appleby:2008tv,Appleby:2009uf}, and it has been argued that the energy density associated with these oscillations will decay in the early Universe, making $\delta R_{\rm osc}$ negligible at late times $z < 10^{3}$. We therefore neglect $\delta R_{\rm osc}$ in what follows. In addition, $\delta R_{\rm ind}$ satisfies $\delta R_{\rm ind} \ll R_{\rm GR}$ in the quasi-static limit, and hence we can also neglect this term. We are left with $R \simeq R_{\rm GR}$, which can now be substituted into the Friedmann equation.  We find 

\begin{equation}\label{eq:e3} {H^{2} \over H_{0}^{2}} = {\Omega_{m0} \over a^{3}} + {\Omega_{\rm r0} \over a^{4}} + \Omega_{\Lambda 0} + {R_{\rm GR} \over 6H_{0}^{2}}\bar{f}_{\rm R} - {\bar{f} \over 6H_{0}^{2}} - a{H_{\rm GR}^{2} \over H_{0}^{2}} R'_{\rm GR}\bar{f}_{\rm RR} , \end{equation}

\noindent where $H_{\rm GR}$ is given by

\begin{equation} {H^{2}_{\rm GR} \over H_{0}^{2}} = {\Omega_{m0} \over a^{3}} + {\Omega_{\rm r0} \over a^{4}} + \Omega_{\Lambda 0} .\end{equation}

\noindent The equation of state of the $f(R)$ effective fluid can be written as

\begin{equation} \label{eq:e4} w \simeq -1 + {a H_{\rm GR} \left( aH_{\rm GR} \bar{f}'_{\rm R}\right)' \over 3H_{0}^{2}\Omega_{\Lambda 0}} - {aH_{\rm GR}^{2} R'_{\rm GR}\bar{f}_{\rm RR} \over 3H_{0}^{2} \Omega_{\Lambda 0}}  . \end{equation}

\noindent The $\bar{f}(R)$ terms on the right hand sides of ($\ref{eq:e3},\ref{eq:e4}$) are of order $H_{\rm GR}^{2}/M^{2} \ll 1$ and constitute small corrections to the standard $\Lambda$CDM expansion.

In the quasi-static limit, the mass of the scalar field is given approximately by 

\begin{equation} M^{2}(a) = {1 \over 3 \bar{f}_{\rm RR} (R_{\rm GR})}  , \end{equation}

\noindent and hence we could in principle write the additional terms in the Friedmann equation solely in terms of $M(a)$. However, this procedure is model dependent, as it requires writing $\bar{f}$ and $\bar{f}_{\rm R}$ in terms of $\bar{f}_{\rm RR}$.

With respect to the perturbations, neglecting all terms of order $H^{2}/M^{2}$ and taking $k\phi /a, k\psi/a \gg \dot{\phi}, \dot{\psi}$, the equations reduce to 

\begin{eqnarray} \label{eq:16} & &  \psi = \left(1 + {2\bar{K}^{2} \over 3 + 2\bar{K}^{2}}\right) \phi , \\ \label{eq:17} & & k^{2}\phi = -4\pi G \left({3 + 2\bar{K}^{2} \over 3 + 3\bar{K}^{2}}\right)a^{2}\rho_{\rm m} \delta_{\rm m} , \\ \label{eq:18}  & & \ddot{\delta}_{\rm m} + 2H \dot{\delta}_{\rm m} - 4\pi G \left({3 + 4\bar{K}^{2} \over 3 + 3\bar{K}^{2}}\right)a^{2}\rho_{\rm m}\delta_{\rm m} = 0 , \end{eqnarray}

\noindent where $\bar{K} \equiv k/aM(a)$. In the quasi-static limit, terms of order $H^{2}/M^{2}$ are small, so we have taken $w = -1$. We are considering modes that satisfy $k \gg aH$ in equations ($\ref{eq:16}-\ref{eq:18}$); superhorizon modes will evolve according to General Relativity as they remain below the scalaron mass scale at all times ($k_{\rm superh} \ll a M(a)$ for all $a$.) 

To summarize, we have argued that the quasi-static approximation, valid whenever $H^{2}/M^{2} \ll 1$, can be used to write the Friedmann equation in its standard General Relativistic form with corrections of order ${\cal O}(H^{2}/M^{2})$. In addition, all modifications to the perturbation equations can be expressed solely in terms of the scalar field mass $M(a)$. In a forthcoming paper, the authors will consider how one might constrain the mass of the scalaron with cosmological data, using the above system of equations.

\subsection{\label{sec:q4}The growth parameter in the quasi static approximation}

Now that we have a dynamical equation for $\delta_{\rm m}$, we can calculate $\gamma$. Performing the same steps as in the General Relativistic case, we find that $\gamma$ is described by the following equation

\begin{eqnarray} \nonumber \gamma'(a,k) = & &  {1 \over \Omega_{\rm m}^{\gamma}\log[\Omega_{\rm m}]}\left[ {3 \over 2}{\Omega_{\rm m} \over a} + \left({3 \over 2}\Omega_{\rm m} - 2\right){\Omega_{\rm m}^{\gamma} \over a} - {\Omega_{\rm m}^{2\gamma} \over a}\right] - {\gamma \over \log[\Omega_{\rm m}]}{3 \over a}\left(\Omega_{\rm m} - 1\right) \\ \label{eq:25} & & \qquad  + {\Omega_{\rm m}^{1- \gamma} \over 2 a \log[\Omega_{\rm m}]} \left( {k^{2} \over  k^{2}  + a^{2}M^{2}}\right) . \end{eqnarray}

\noindent This expression reduces to the General Relativistic equation ($\ref{eq:5}$) (with $w = -1$) for $M(a) \to \infty$, as expected. The effect of modifying gravity is to introduce the last term on the right hand side of ($\ref{eq:25}$).

To obtain a solution valid for $z > 1$ we expand $\Omega_{\rm m} \simeq 1 - \Omega_{\Lambda 0}a^{3}/\Omega_{\rm m0} + {\cal O}(a^{6})$, in which case equation ($\ref{eq:25}$) reduces to 

\begin{equation}\label{eq:51} \gamma'(a,k) = -{11 \over 2a}\gamma(a,k) + {3 \over a} - {1 \over 2\omega_{0}a^{4}}{k^{2}  \over k^{2}  + a^{2}M^{2}(a)} , \end{equation}

\noindent which has solution

\begin{equation}\label{eq:50} \gamma(a,k) = {6 \over 11} + \gamma_{0}(k)a^{-11/2} - {k^{2} \over 2\omega_{0}}a^{-11/2}\int_{0}^{a}{\bar{a}^{3/2} d\bar{a} \over k^{2} + \bar{a}^{2}M^{2}(\bar{a})} , \end{equation}

\noindent where $\omega_{0} \equiv \Omega_{\Lambda 0}/\Omega_{\rm m0}$ and $\gamma_{0}$ is an arbitrary function of $k$, which is set to zero in order to recover the General Relativistic limit as $a \to 0$. We note that in deriving ($\ref{eq:51}$), we did not expand the term $k^{2}/(k^{2}+a^{2}M^{2}(a))$ in $a \ll 1$, as it is a non-linear function of the scale factor \footnote{See \cite{Linder:2007hg} for earlier work on integral representations of $\gamma$ in modified gravity models.}.

For generic $f(R)$ functions, the integral in ($\ref{eq:50}$) does not admit an analytic solution, however from ($\ref{eq:51}$) we can broadly describe the evolution of $\gamma$ for this class of models. At early times $z \gtrsim 10$ the mass of the Scalaron will satisfy $aM(a) \gg k$ for all modes of interest, and deviations from General Relativity will be suppressed. However, whenever any given $k$ mode crosses the scalaron horizon $k = a_{\rm c}M(a_{\rm c})$, the last term on the right hand side of ($\ref{eq:51}$) will dominate and $\gamma$ will decrease. There will be a zero in $\gamma'$ due to the fact that the first and second terms grow relative to the third, and $\gamma$ will increase to the present, with asymptotic behaviour $\gamma \to 2/3$ for $a \gg 1$ (this can be derived from the full equation ($\ref{eq:25}$).)

Equation ($\ref{eq:51}$) and approximate solution ($\ref{eq:50}$) have been constructed by expanding the full equation ($\ref{eq:25}$) for $a \ll1$. Whilst this power series is strongly convergent and hence valid up to $a \simeq 1/2$, we cannot obtain a solution in the region of particular interest (that is, $z < 1$) using this approach. However, as we shall see in section \ref{sec:i7} our approximation is sufficient to accurately calculate the matter power spectrum at all redshifts.

\subsection{\label{sec:l6}Application to specific $f(R)$ models}

Now that we have an approximate, model independent form for $\gamma$ we can apply it to some existing $f(R)$ functions in the literature. Specifically, we consider the functional forms ($\ref{eq:p1},\ref{eq:p2}$). Our approach will be to solve the perturbation equation ($\ref{eq:18}$) numerically for $\delta_{\rm m}$, and use this in the definition ($\ref{eq:2}$) to obtain $\gamma$. We then compare this to our approximate $\gamma$, obtained by numerically integrating ($\ref{eq:50}$).

\begin{figure}[!h]
\centering
\mbox{\resizebox{0.8\textwidth}{!}{\includegraphics[angle=0]{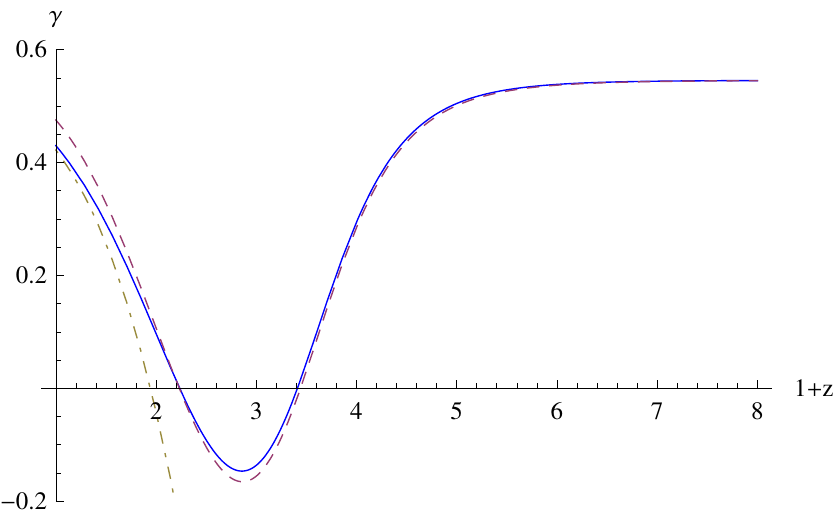}}}
\caption{\label{fig:1} The growth parameter $\gamma$ for the model ($\ref{eq:p2}$), with model parameters $n=3$, $\lambda = 5\times 10^{-4}$ (solid), where we have taken $k = 0.14h {\rm Mpc}^{-1}$. We have also exhibited approximate expressions for $\gamma$ derived in section \ref{sec:2} in the $a \ll 1$ limit (dashed) and $a \sim 1$ (dot-dashed) limits. We observe a close agreement at all redshifts.}
\end{figure}

\begin{figure}[!h]
\centering
\mbox{\resizebox{0.8\textwidth}{!}{\includegraphics[angle=0]{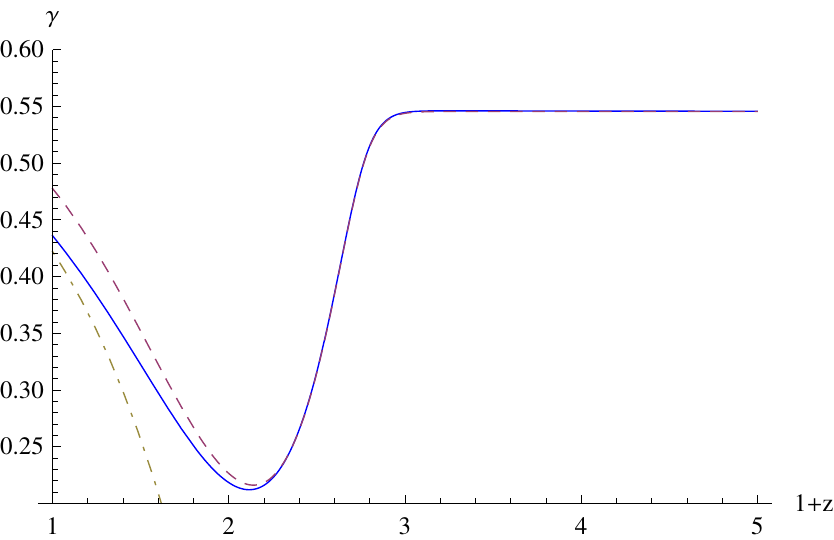}}}
\caption{\label{fig:3} The growth parameter $\gamma$ for the model ($\ref{eq:p1}$), with model parameters $b=2$, $g = 0.45$ (solid), where we have taken $k = 0.14h {\rm Mpc}^{-1}$. We have also exhibited approximate expressions for $\gamma$ derived in section \ref{sec:2} in the $a \ll 1$ limit (dashed) and $a \sim 1$ (dot-dashed) limits.}
\end{figure}

Taking an initial redshift $z_{\rm i} = 100$ and using model parameters $n = 3/2$, $\lambda = 5\times 10^{-4}$, $b = 2$, $g = 0.45$, the results are exhibited in Figs.\ref{fig:1},\ref{fig:3}.  We have neglected radiation, anticipating the fact that any deviations from General Relativity will occur for $z < 10$. We observe a close agreement between analytic and numerical solutions for $z > 1$, and a loss of accuracy in our approximation for $z <1$, as expected. 

The dot-dashed curve in Figs.\ref{fig:1},\ref{fig:3} is an approximate analytic solution for $\gamma$ in the $a \sim 1$ region, which is discussed in the Appendix. For the power law model, we observe a close agreement between the numerical and analytic solutions at low redshift, however for the exponential model our approximation for $z < 1$ is not particularly accurate. This is due to the fact that $k$ modes of interest to the linear matter power spectrum only deviate from General Relativity at low redshift $z \lesssim 1$ for this model, whereas the key assumption made in obtaining this curve is that modifications to GR occur at $z > 1$. We direct the reader to the Appendix for further details.

\begin{figure}[!h]
\centering
\mbox{\resizebox{0.83\textwidth}{!}{\includegraphics[angle=0]{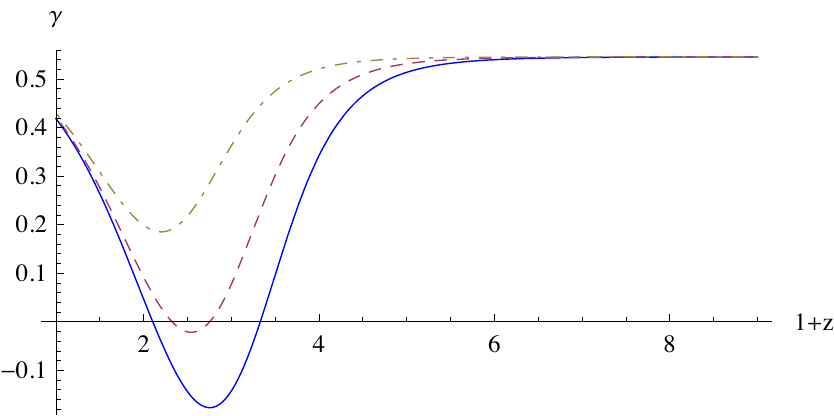}}}
\caption{\label{fig:6} The growth parameter $\gamma(a,k)$ for the model ($\ref{eq:p2}$), taking model parameters as in figs.\ref{fig:1} for $k= 0.15h{\rm Mpc^{-1}}$ (solid), $k = 0.1 h{\rm Mpc^{-1}}$ (dashed), $k=0.05 h {\rm Mpc^{-1}}$. (dot-dashed). }
\end{figure}

\begin{figure}[!h]
\centering
\mbox{\resizebox{0.8\textwidth}{!}{\includegraphics[angle=0]{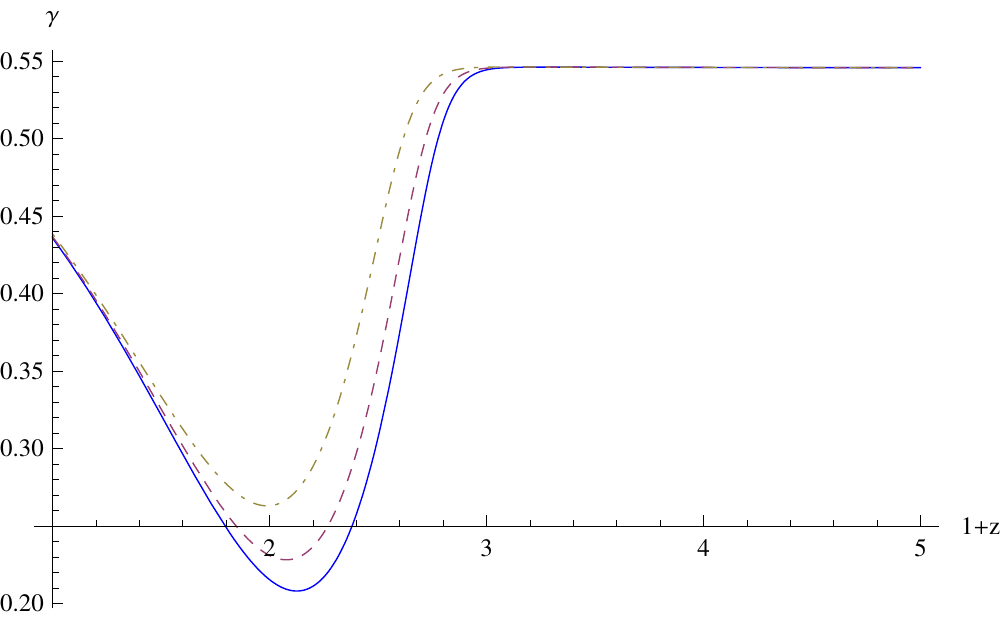}}}
\caption{\label{fig:7} The growth parameter $\gamma(a,k)$ for the model ($\ref{eq:p1}$), taking model parameters as in figs.\ref{fig:1} for $k= 0.15h{\rm Mpc^{-1}}$ (solid), $k = 0.1 h{\rm Mpc^{-1}}$ (dashed), $k=0.05 h {\rm Mpc^{-1}}$. (dot-dashed).  }
\end{figure}

In Figs.\ref{fig:6},\ref{fig:7} we show $\gamma(a,k)$ for three different modes; $k=0.15,0.1,0.05 h{\rm Mpc}^{-1}$. For the model parameters chosen, $\gamma$ exhibits only a very weak $k$ dependence at late times, taking the value $\gamma \simeq 0.43$. This value has been obtained previously in the literature for a number of different models (see for example \cite{Tsujikawa:2009ku}), and we derive it explicitly in the Appendix.

\section{\label{sec:i7}Power spectrum reconstruction from $\gamma$}

\begin{figure}[!h]
\centering
\mbox{\resizebox{0.8\textwidth}{!}{\includegraphics[angle=0]{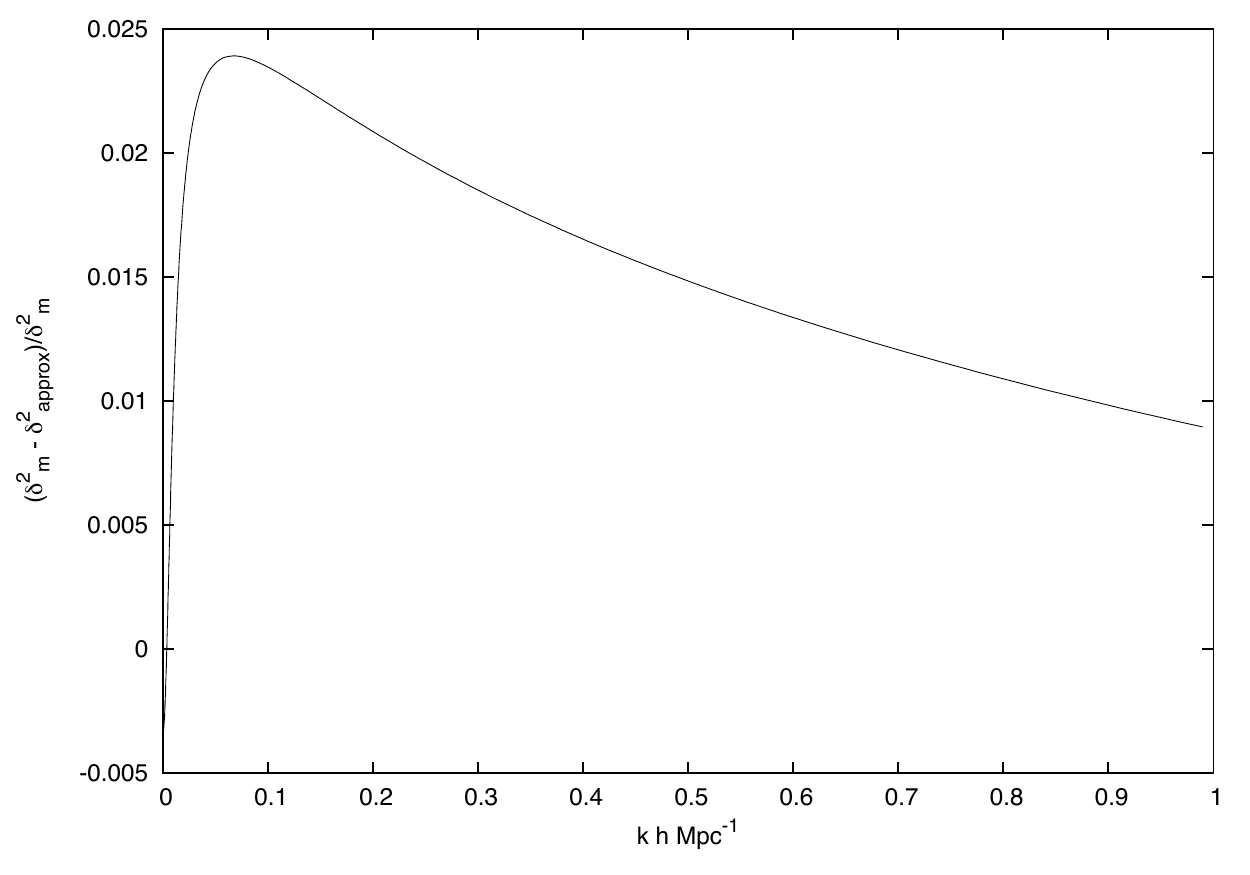}}}
\caption{\label{fig:12} The quantity $(\delta_{\rm m}^{2} - \delta_{\rm approx}^{2})/\delta_{\rm m}^{2}$, which is the fractional difference in $\delta_{\rm m}^{2}$ using the two different approaches explained in the text. The fractional error is small (less than $\sim 2 \%$), suggesting that our expression ($\ref{eq:50}$) for $\gamma$ is suitable to reconstruct the power spectrum.  }
\end{figure}

\begin{figure}[!h]
\centering
\mbox{\resizebox{0.8\textwidth}{!}{\includegraphics[angle=0]{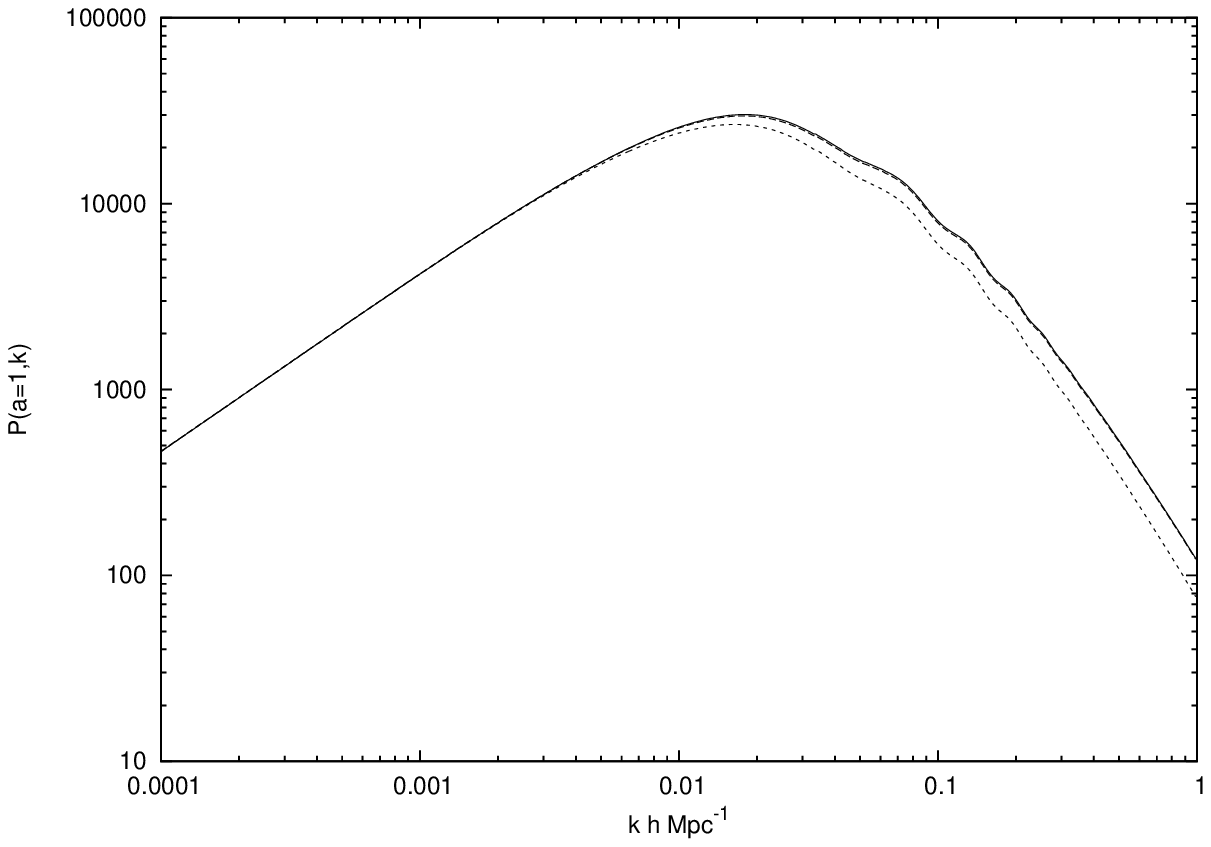}}}
\caption{\label{fig:11} Power spectrum for the model ($\ref{eq:p2}$), with model parameters $n = 1$, $\lambda = 1 \times 10^{-3}$ (solid line) and the General Relativistic power spectrum (dotted line). We observe an increased power at small scales, as expected. The dashed line, practically indistinguishable from the solid line, is the power spectrum obtained using our approximate scheme (that is, numerically integrating ($\ref{eq:50}$) and using the resulting $\gamma$ to obtain $\delta_{\rm approx}$ from ($\ref{eq:i10}$).   }
\end{figure}

Whilst our approximate form for $\gamma$, equation ($\ref{eq:50}$), is only accurate to $\sim 10\%$ at low redshift,  we now use it to obtain the power spectrum $P(a=1,k)$, and compare our result to the power spectrum obtained by numerically evolving the full perturbation equations ($\ref{eq:16}-\ref{eq:18}$). Taking the power law $f(R)$ function ($\ref{eq:p2}$) and the model parameters $n =1$, $\lambda = 10^{-3}$,  our results are exhibited in Figs.\ref{fig:12},\ref{fig:11}. In fig.\ref{fig:12} we exhibit the difference $\delta^{2}_{\rm approx}(a=1,k)/\delta^{2}_{\rm m}(a=1,k) - 1$, where $\delta_{\rm m}$ is the density perturbation obtained by numerically evolving  ($\ref{eq:18}$) and $\delta_{\rm approx}$ is obtained by integrating the expression

\begin{equation} \label{eq:i10} \delta_{\rm approx} = \exp \left[ \int \Omega_{\rm m}^{\gamma} d\ln a \right] , \end{equation}

\noindent using our approximate solution ($\ref{eq:50}$) for $\gamma$. We observe a close agreement between the two approaches, which suggests that our approximation ($\ref{eq:50}$) can be used to obtain the power spectrum for these models at any redshift, despite the apparent loss of accuracy at $z \sim 0$. We have confirmed this statement for a wide range of parameter choices. 

In fig.\ref{fig:11} we exhibit the power spectrum for the $f(R)$ model ($\ref{eq:p2}$) at $z = 0$. This has been obtained by evolving ($\ref{eq:18}$) in the range $z = (20,0)$ to obtain $\delta_{\rm m}(a,k)$, and normalizing the power spectrum to its General Relativistic value at $z=20$. The resulting $f(R)$ and General Relativistic power spectra are exhibited as solid and dotted lines respectively. We observe an increased power on small scales for the modified gravity model. This is due to the fact that for the model parameters chosen, high $k$ modes will satisfy $k > aM(a)$ at late times. Whenever this occurs, there will be an order $\sim {\cal O}(10\%)$ increase in $G_{\rm eff}$ in the $\delta_{\rm m}$ dynamical equation, leading to enhanced power. Low $k$ modes, which do not satisfy $k > aM(a)$ at any redshift, will not observe any increase in $G_{\rm eff}$ and will evolve according to General Relativity.

In fig.\ref{fig:11} we have exhibited the $f(R)$ power spectrum using both the full equation ($\ref{eq:18}$) to obtain $\delta_{\rm m}$ and the approximate solution $\delta_{\rm approx}$; the approximate solution is the dashed curve which closely overlaps the full solution in fig.\ref{fig:11}. We observe an excellent agreement between the two methods.

\subsection{\label{sec:7}Comment on the $\gamma$ parameterization for $f(R)$ gravity}

So far in this paper we have derived the expression ($\ref{eq:50}$) for the growth parameter $\gamma$, involving a model dependent integral over the mass of the scalaron. The $M(a)$ dependence significantly alters the behaviour of $\gamma$ from its General Relativistic form; $\gamma$ will now be a non-analytic function of $a$, $k$, $\Omega_{\rm m0}$, $H_{0}$ and will depend on the explicit functional form of $f(R)$. To obtain $\delta_{\rm m}$ using $\gamma$, one must numerically integrate  the expression ($\ref{eq:j1}$) relating $\gamma$ and $\delta_{\rm m}$, where $\gamma$ itself involves a non-analytic, model dependent integral over $M(a)$. Given these facts, it is not obvious why one should persevere with $\gamma$ as a parameterization for growth in $f(R)$ gravity.

In addition, one should be careful not to construct an approximate functional form for $\gamma$, valid over a particular redshift range. For example, in the previous section we found that $\gamma$ will exhibit only a very weak $k$ dependence at low redshift. One might therefore be tempted to construct a simple functional form for $\gamma = \gamma(a)$ for $a \sim 1$ and use this in ($\ref{eq:j1}$) to obtain $\delta_{\rm m}$. However, this approach will not uniquely specify the power spectrum, as the $k$ dependence of $\delta_{\rm m}$ will remain undetermined. This is due to the fact that $\gamma$ is only defined via the ratio $\delta'_{\rm m}/\delta_{\rm m}$ (see ($\ref{eq:2}$)), and so writing $\delta_{\rm m}$ as $\delta_{\rm m} = \delta_{0}(k)\delta_{1}(a,k)$ (where $\delta_{1}(a,k)$ is only weakly $k$ dependent at $z \sim 0$), it is clear that $\gamma$ will give us no information on $\delta_{0}(k)$. In this respect, one can think of $f(R)$ models as modifying the power spectrum at late times in two ways; changing the time evolution of the amplitude and introducing an additional component to the transfer function. Phenomenological constructions of $\gamma(a)$ will yield information on the time evolution, but not the modified transfer function. Care should be taken for any model in which $\delta_{\rm m}$ exhibits both scale and time dependence.

\section{Parameterizing the scalaron mass $M(a)$}

We have argued above that directly parameterizing $\gamma(a,k)$ is problematic in $f(R)$ gravity. Given that we can write the modified Friedmann and perturbation equations in terms of the scalaron mass $M(a)$, it seems more appropriate to parameterize this quantity. Once we have a functional form for $M(a)$, one can then use it in ($\ref{eq:50}$) to obtain an expression for $\gamma$, if one wishes to persevere with this parameterization. In doing so, both the $a$ and $k$ dependence of $\gamma$ and hence the power spectrum will be accounted for.

\begin{figure}
\centering
\mbox{\resizebox{0.6\textwidth}{!}{\includegraphics[angle=0]{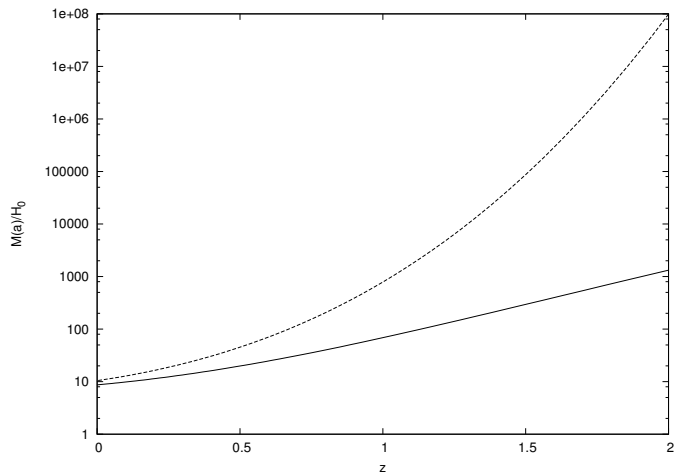}}}
\caption{\label{fig:mp1} The scalar field mass $M(a)/H_{0}$ for the exponential (dashed) and power law (solid) models ($\ref{eq:p1},\ref{eq:p2}$), with model parameters $n = 3$, $\lambda = 2\times 10^{-3}$, $g = 0.48$, $b = 3$. Both monotonically increase to the past. }
\end{figure}

The advantage of using $M(a)$ is that it is not scale dependent, and the essential features of viable $f(R)$ models can be captured with a very simple functional form. All we require is a function $M(a)$ which monotonically grows to the past, such that $R_{\rm GR}/M^{2}(a) \to 0$ as $a \to 0$ (one should introduce a cutoff to ensure that $M(a)$ does not diverge as $a \to 0$, however this is unnecessary as we only consider the late universe $z \lesssim1000$.) In fig.\ref{fig:mp1} we show the simple behaviour of the mass $M(a)/H_{0}$ for the power law and exponential models, for parameter choices $n = 3$, $\lambda = 2\times 10^{-3}$, $g = 0.48$, $b = 3$ and $\Omega_{\rm m0} = 0.3$. 

A simple functional form for $M(a)$, which is representative of `viable' $f(R)$ models, is given by

\begin{equation} \label{eq:p10} M(a) = {\mu H_{0} \over a^{\sigma}} , \end{equation}

\noindent where $\mu > 1$ and $\sigma > 4$ are free parameters. This mass corresponds to an $f(R)$ model that is  is free from instabilities for $ 0 \leq z \leq 10^{3}$, and reduces to General Relativity at early times (subject to $\sigma > 4$; this parameter dictates how quickly the mass diverges to the past.) It therefore constitutes a viable $f(R)$ model in the regime of interest. We note that this parameterization has been considered previously; see \cite{Zhao:2008bn}.

Using this mass in ($\ref{eq:50}$), we can write the growth parameter $\gamma$ in terms of a hypergeometric function ${}_{2}F_{1}[a,b,c;x]$,

\begin{equation} \label{eq:l1} \gamma  = {6 \over 11} - {k^{2} \over M^{2}(a)} {\Omega_{\rm m0} \over a^{5}\Omega_{\Lambda 0}(1+4\sigma)} \hspace{2mm} {}_{2}F_{1} \left[ 1,{4\sigma+1 \over 4\sigma-4};{8\sigma-3 \over 4\sigma-4}; - { k^{2} \over a^{2}M^{2}(a)}\right] . \end{equation}

\noindent Similarly, we can write the modified Friedmann equation ($\ref{eq:e3}$) solely in terms of $M(a)$,

\begin{equation}\label{eq:l2} {H^{2} \over H_{0}^{2}} = {\Omega_{\rm m0} \over a^{3}} + \Omega_{\Lambda 0} + \alpha(a){H_{0}^{2} \over M^{2}(a)} , \end{equation}

\noindent where the function $\alpha(a)$ is given by

\begin{equation} \alpha(a) = {3\Omega_{\rm m0} \over 4a^{6}} \left[ 4\Omega_{\Lambda 0} a^{3} {2\sigma - 5 \over 2\sigma - 3} + \Omega_{\rm m0} {4\sigma - 13 \over \sigma - 3} \right] . \end{equation}

\noindent By using ($\ref{eq:l1}$) in ($\ref{eq:i10}$), numerically integrating this expression over low redshifts $z < 20$ and normalising the power spectrum to General Relativity at $z = 20$, one can obtain the linear matter power spectrum for this parameterization. Hence ($\ref{eq:l1}$) and ($\ref{eq:l2}$) are sufficient to fully parameterize the growth and expansion histories for the class of `viable' $f(R)$ models, subject to the quasi static approximation $M(a) > H(a)$. 

\begin{figure}
\centering
\mbox{\resizebox{0.6\textwidth}{!}{\includegraphics[angle=0]{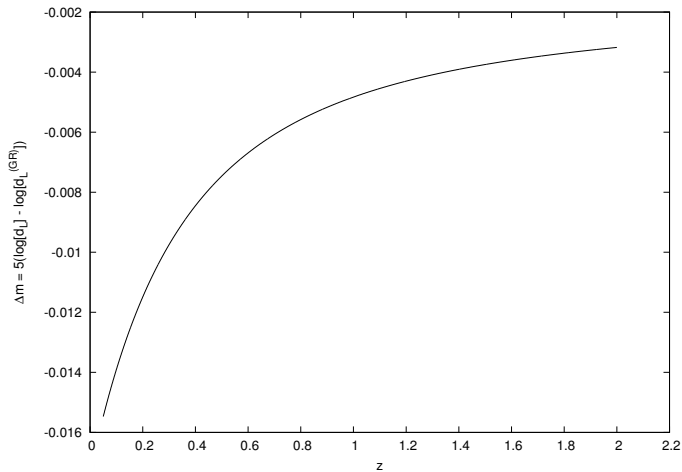}}}
\caption{\label{fig:m1} The difference between the luminosity distance for the modified gravity model characterized by the mass ($\ref{eq:p10}$) and the General Relativistic luminosity distance $d_{\rm L}^{\rm (GR)}$. There is practically no modified gravity signal in $d_{\rm L}$, indicating that the expansion history will be an ineffective probe of these models.}
\end{figure}

\begin{figure}
\centering
\mbox{\resizebox{0.6\textwidth}{!}{\includegraphics[angle=0]{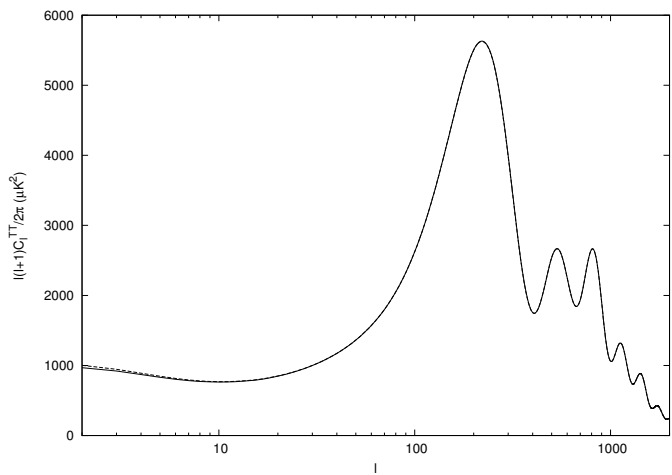}}}
\caption{\label{fig:m2} The CMB angular power spectrum for the model ($\ref{eq:p10}$) (solid line) and the General Relativistic power spectrum (dashed line), taking $\Omega_{\rm m0} = 0.29$, $\Omega_{\Lambda 0} = 0.71$, $n_{\rm s} = 1$, $h = 0.7$. We observe no significant modified gravity signature, even at large angular scales.  }
\end{figure}

\begin{figure}
\centering
\mbox{\resizebox{0.6\textwidth}{!}{\includegraphics[angle=0]{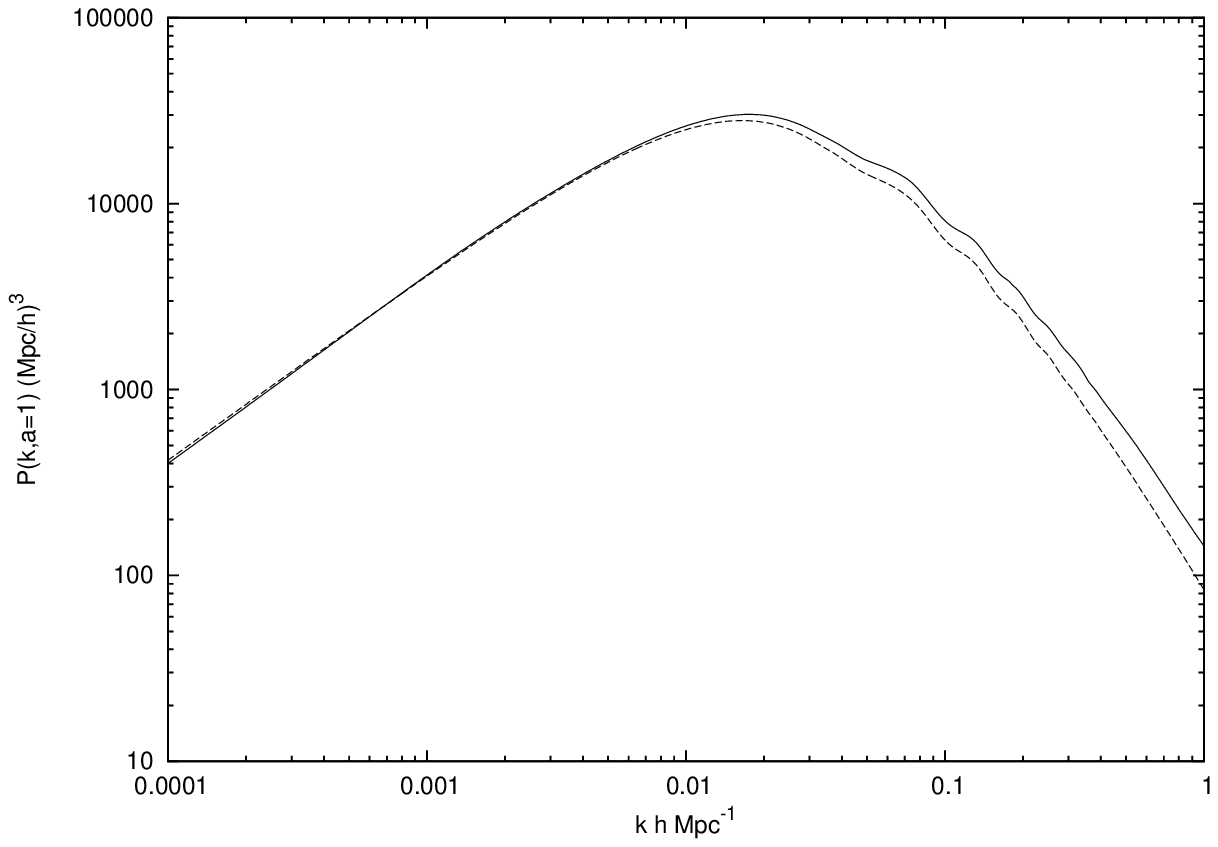}}}
\caption{\label{fig:m3} The matter power spectrum for the model ($\ref{eq:p10}$) (solid line) and the General Relativistic power spectrum (dashed line), $\Omega_{\rm m0} = 0.29$, $\Omega_{\Lambda 0} = 0.71$, $n_{\rm s} = 1$, $h = 0.7$. We observe a significant increase in power at small scales, as expected. }
\end{figure}

To highlight the effect of $M(a)$ on various observational probes, we exhibit the luminosity distance, the CMB angular power spectrum and the matter power spectrum for the model ($\ref{eq:p10}$) in figs.\ref{fig:m1}-\ref{fig:m3}, taking model parameters $\mu = 6$, $\sigma=4$. Specifically, in fig.\ref{fig:m1} we show the difference $\triangle m = 5(\log[d_{\rm L}] - \log[d_{\rm L}^{\rm (GR)}])$ between the luminosity distance for the modified gravity model ($\ref{eq:p10}$) and its $\Lambda$CDM value $d_{\rm L}^{\rm (GR)}$, where $d_{\rm L}$ is defined in the usual fashion

\begin{equation} d_{\rm L} (z) = (1+z) \int_{0}^{z} {dz' \over H(z')} .\end{equation}

\noindent We observe practically no modified gravity signal in either the luminosity distance or the CMB angular power spectrum, indicating that the most stringent constraints on these models will be be derived from the matter power spectrum, as shown in Fig.\ref{fig:m3}.

\section{Discussion}

In this work we have considered the evolution of density perturbations for the class of so-called `viable' $f(R)$ modified gravity models. By using the quasi static approximation, we have written the Friedmann equation as an expansion around General Relativity, with corrections of order $H^{2}_{\rm GR}/M^{2}$. The perturbation equations can similarly be written in terms of the mass of the scalar field, and hence we can parameterize both the expansion and growth histories solely in terms of $M(a)$. In a future publication, we will use the modified Friedmann and perturbation equations to derive constraints on the mass of the scalar field in $f(R)$ models with cosmological probes.

We have also constructed an approximate functional form for the growth parameter for this class of models, again in terms of the mass $M(a)$. Whilst our approximate solution is only accurate to $\sim 10\%$ for $0 <z <1$, it can still be used to accurately reconstruct the power spectrum at all redshifts. We have shown that by specifying a suitable mass function $M(a)$, equations ($\ref{eq:e3}$) and ($\ref{eq:50}$) are sufficient to constrain this class of models with cosmological data.

We would like to conclude by comparing our approach to existing parameterizations in the literature. The perturbation equations ($\ref{eq:16}-\ref{eq:18}$) belong to a parameterization considered in \cite{Amendola:2007rr},\cite{Jain:2007yk}, where arbitrary functions $\eta(a,k)$ and $\mu(a,k)$ were introduced into the perturbation equations

\begin{eqnarray} \label{eq:d1} & & \psi = \left(1 + \eta(a,k)\right) \phi , \\  \label{eq:d2} & & k^{2}\phi = -4\pi G\mu(a,k) a^{2}\rho_{\rm m} \delta_{\rm m}  . \end{eqnarray} 

\noindent Equations ($\ref{eq:d1},\ref{eq:d2}$) have a broad range of applicability and represent an extremely general parameterization of modified gravity models, of which we have only considered a very specific subset. However, one advantage to our approach is that we have an action from which to derive the field equations, and hence both the background expansion and perturbation equations are consistent (it is not clear how the background expansion is modified if we simply choose $\eta(a,k)$ and $\mu(a,k)$.) We also note that when equations ($\ref{eq:d1},\ref{eq:d2}$) are implemented, typically the $k$ dependence of $\eta$ and $\mu$ is neglected.

Finally, we comment on a series of papers containing the most comprehensive cosmological constraints on $f(R)$ models to date \cite{Song:2006ej},\cite{Hu:2007pj},\cite{Lombriser:2010mp}. In these works, the $f(R)$ terms in the modified Friedmann equation ($\ref{eq:i5}$) are written as the energy density of a fluid (typically a perfect fluid with constant equation of state), and this equation is treated as a second order differential equation for $f(a)$. By demanding that the model reduces to General Relativity as $a \to 0$, one of the integration constants is fixed. There is only one remaining free parameter, taken to be $B_{0}$ as defined in the introduction. Then the perturbation equations in all three domains of interest (superhorizon, subhorizon and non-linear) are written in terms of $B(a)$, and $B_{0}$ is constrained using cosmological data sets. This approach is more sophisticated than ours, in the sense that the non-linear regime has also been considered (we will analyse the non-linear regime in a forthcoming publication, see also \cite{Koyama:2009me}.)

The most significant difference between this approach and ours is that we have not fixed the background expansion history; although it remains close to $\Lambda$CDM there will be corrections of order $H_{\rm GR}^{2}/M^{2}$. The mass of the scalar field $M(a)$ therefore contains two parameters. One dictates the value of $M(a)$ relative to the Hubble parameter at the present time (this is essentially $B_{0}$), and the other dictates how quickly $M(a)$ grows to the past. By fixing the evolution of $H(a)$, the time evolution of $M(a)$ is fixed, and it can be shown that an $f(R)$ model with equation of state $w = -1$ corresponds to an $f(R)$ model given by

\begin{equation}\label{eq:kj10}  f(R) \simeq -{R_{\rm vac} \over 2} + A R_{\rm vac} \left({R_{\rm vac} \over R}\right)^{p_{+}/3} , \end{equation}

\noindent during the matter dominated epoch $1< z  < 10^{3}$, where $A$ is the free model parameter (which is related to $B_{0}$), and $p_{+} = (-7+\sqrt{73})/4 \simeq 0.4$. This is an $f(R)$ model with particularly slow asymptotic behaviour to the past; it is of the form ($\ref{eq:p2}$) with $n\simeq 0.06$. It would be instructive to see what constraints on $B_{0}$ can be obtained for models where $M(a)$ possesses a steeper redshift dependence. This will be the subject of a forthcoming publication.

\vspace{5mm}

{\em Acknowledgments: The authors would like to thank Eric Linder and Shaun Thomas for helpful conversations. SA would like to thank Scott Daniel and Eric Linder for discussions that have greatly assisted in modifying CAMB; see \cite{Daniel:2010yt} for a more comprehensive parameterization not specific to $f(R)$ gravity.}

\section{\label{sec:3}Appendix: low redshift behaviour of $\gamma$}

To obtain an approximate expression for $\gamma$ in the region $0 < z < 1$,  we use the simplifying assumption that $\theta = (3 + 4\bar{K}^{2})/(3 + 3\bar{K}^{2})$ in ($\ref{eq:18}$) acts as a heaviside function, interpolating between $\theta(a,k) \simeq 1$ for $k \ll a M(a)$ and $\theta \simeq 4/3$ for $k \gg a M(a)$.  We therefore solve the equations

\begin{eqnarray} & & a^{2} \delta''_{\rm m 1}(a,k) + 3a \left[ 1 - {\Omega \over 2}\right]\delta'_{\rm m 1}(a,k) = {3 \over 2} \Omega  \delta_{\rm m 1} , \\ & & a^{2} \delta''_{\rm m 2}(a,k) + 3a \left[ 1 - {\Omega \over 2}\right]\delta'_{\rm m 2}(a,k) = 2 \Omega \delta_{\rm m 2} , \end{eqnarray}

\noindent and match $\delta_{\rm m 1,2}$ and $\delta'_{\rm m 1,2}$ at $a = a_{\rm c}(k)$. If we assume that the relevant modes cross the scalaron horizon for $z > 1$, where the small $a$ expansion can be used to match the two solutions at $a = a_{\rm c}$, we obtain the following solution for $z < 1$

\small
\begin{eqnarray} \label{eq:iop20}  & & \delta_{\rm m} \simeq  -\delta_{0}  \left({7 + \sqrt{33} \over 8}\right)    \left({\Omega_{\rm m0} \over \Omega_{\rm de0}}\right)^{m_{+}/3} a_{\rm c}^{1-m_{+}}(1-\Omega_{\rm m})^{(-1+\sqrt{33})/12} \times \\  \nonumber & & \left( C_{1} {}_{2}F_{1}\left[{-1+\sqrt{33} \over 12},{5+\sqrt{33} \over 12},{1 \over 3},\Omega_{\rm m}[a]\right]  + \Omega_{\rm m}^{2/3} C_{2} {}_{2}F_{1}\left[{7+\sqrt{33} \over 12},{13+\sqrt{33} \over 12},{5 \over 3},\Omega_{\rm m}[a]\right]  \right) ,  \end{eqnarray}

\normalsize 

\noindent where the constants $C_{1}$ and $C_{2}$ are given by

\begin{eqnarray} & & B_{1}^{(+)} = { \Gamma[1/3]\Gamma[\sqrt{33}/6] \over \Gamma[(5 + \sqrt{33})/12] \Gamma[(-1+\sqrt{33})/12]} , \\ & & B_{1}^{(-)} = { \Gamma[1/3]\Gamma[-\sqrt{33}/6] \over \Gamma[(5 - \sqrt{33})/12] \Gamma[(-1-\sqrt{33})/12]} , \\& & B_{2}^{(+)} = { \Gamma[5/3]\Gamma[\sqrt{33}/6] \over \Gamma[(13 + \sqrt{33})/12] \Gamma[(7+\sqrt{33})/12]} ,  \\ & & B_{2}^{(-)} = { \Gamma[5/3]\Gamma[-\sqrt{33}/6] \over \Gamma[(13 - \sqrt{33})/12] \Gamma[(7-\sqrt{33})/12]} , \\ & & C_{1} = {B_{2}^{(+)} \over B_{1}^{(+)}\left(B_{2}^{(-)} + {9 +\sqrt{33} \over 18}B_{2}^{(+)}\right)- B_{2}^{(+)}\left(B_{1}^{(-)} + {9 +\sqrt{33} \over 18}B_{1}^{(+)}\right)}  , \\ & & C_{2} = {B_{1}^{(+)} \over B_{2}^{(+)}\left(B_{1}^{(-)} + {9 +\sqrt{33} \over 18}B_{1}^{(+)}\right)- B_{1}^{(+)}\left(B_{2}^{(-)} + {9 +\sqrt{33} \over 18}B_{2}^{(+)}\right)} , \end{eqnarray}

\noindent and $m_{+}$ is given by

\begin{equation} m_{+} = {-1 + \sqrt{33} \over 4} . \end{equation}

\noindent We stress that ($\ref{eq:iop20}$) is only valid at late times, and only for $k$ modes that have crossed the `scalaron horizon' at early times $z > 1$. The key approximation that has been used in deriving our solution; that $\theta(a,k)$ can be approximated by a discontinuous step function, is not applicable for modes that cross the horizon for $z \lesssim 1$. 

The solution ($\ref{eq:iop20}$)  can be written as $\delta_{\rm m} = \delta_{0} B(a) a_{\rm c}^{1-m_{+}}$, where $B(a)$ is a model independent function of $a$. All $k$ and $f(R)$ model dependence is incorporated in the $a_{\rm c}^{1-m_{+}}$ term, which will introduce a tilt to the matter power spectrum, relative to the standard $\Lambda$CDM case. This tilt can be derived explicitly, by inverting the expression $a_{\rm c} M(a_{\rm c}) = k$ for any given model, and substituting the resulting $a_{\rm c}(k)$ into our expression for $\delta_{\rm m}$. For the power law model ($\ref{eq:p2}$) we find $\delta_{\rm m} = A_{0} B(a) k^{(\sqrt{33}-5)/(12n+8)}$, where $A_{0}$ is an unimportant constant. This power law $k$ dependence has been obtained in \cite{Motohashi:2010qj,Motohashi:2010tb,Tsujikawa:2009ku}; for the exponential model ($\ref{eq:p1}$), $\delta_{\rm m}$ acquires a logarithmic dependence.

Our calculation is in agreement with existing work in the literature. Since $\delta_{\rm m}$ is a separable function of $k$ and $a$ for $z \sim 0$, it follows that $\gamma$ is independent of $k$ for $z \sim 0$, at the level of  approximation to which we are working. Using ($\ref{eq:iop20}$) in the definition of $\gamma$ and taking $\Omega_{\rm m0} = 0.27$, we obtain $\gamma|_{a=1} \simeq 0.42$. Both of these results are in agreement with \cite{Tsujikawa:2009ku}, where it was noted that at $z = 0$, $\gamma$ has a small dispersion in $k$ and takes the value $\gamma \simeq 0.4-0.43$.

\end{document}